\documentclass[12pt]{article}
\usepackage{amsthm,amssymb,amsmath,url,amsfonts,mathrsfs}
\title{Long-Time Behavior of Macroscopic Quantum Systems: Commentary Accompanying the English Translation of John von Neumann's 1929 Article on the Quantum Ergodic Theorem}
\author{
Sheldon Goldstein,\footnote{Departments of Mathematics and Physics, 
	Rutgers University, 
	110 Frelinghuysen Road, Piscataway, NJ 08854-8019, USA.
	E-mail: oldstein@math.rutgers.edu}\ \ 
Joel L. Lebowitz,\footnote{Departments of Mathematics and Physics, 
	Rutgers University, 
	110 Frelinghuysen Road, Piscataway, NJ 08854-8019, USA.
	E-mail: 
	lebowitz@math.rutgers.edu}\\ 
Roderich Tumulka,\footnote{Department of Mathematics, Rutgers University, 
	110 Frelinghuysen Road, Piscataway, NJ 08854-8019, USA. E-mail: 
	tumulka@math.rutgers.edu}{}\ \ and
Nino Zangh\`\i\footnote{Dipartimento di Fisica dell'Universit\`a di
	Genova and INFN sezione di Genova, Via Dodecaneso 33, 16146
	Genova, Italy. E-mail: 
	zanghi@ge.infn.it}
}
\date{June 30, 2010}

\theoremstyle{plain}
\newtheorem{thm}{Theorem}
\newtheorem{claim}{Claim}
\newtheorem{defn}{Definition}

\newcommand{\ket}[1]{\vert#1\rangle}
\newcommand{\bra}[1]{\langle#1\vert}
\newcommand{\pr}[1]{\ket{#1}\bra{#1}}
\DeclareMathOperator{\tr}{tr}
\newcommand{\Ddelta}{\boldsymbol{\mathsf{\Delta}}}
\newcommand{\Aa}{\boldsymbol{\mathsf{A}}}
\newcommand{\Bb}{\boldsymbol{\mathsf{B}}}
\newcommand{\Ee}{\boldsymbol{\mathsf{E}}}
\newcommand{\Hh}{\boldsymbol{\mathsf{H}}}
\newcommand{\Qq}{\boldsymbol{\mathsf{Q}}}
\newcommand{\Pp}{\boldsymbol{\mathsf{P}}}
\newcommand{\Uu}{\boldsymbol{\mathsf{U}}}
\newcommand{\M}{\mathfrak{M}}

\newcommand{\RRR}{\mathbb{R}}
\newcommand{\EEE}{\mathbb{E}}

\newcommand{\scp}[2]{\langle #1| #2 \rangle}
\newcommand{\Hilbert}{\mathscr{H}}
\newcommand{\be}{\begin{equation}}
\newcommand{\ee}{\end{equation}}

\newcommand{\D}{D} 
\newcommand{\dd}{d} 
\newcommand{\decomp}{\mathscr{D}} 
\newcommand{\vq}{\boldsymbol{q}}
\newcommand{\vp}{\boldsymbol{p}}
\newcommand{\A}{\mathscr{A}}
\newcommand{\sphere}{\mathbb{S}}
\newcommand{\total}{\mathrm{total}}
\newcommand{\mc}{\mathrm{mc}}
\newcommand{\can}{\mathrm{can}}
\newcommand{\eq}{\mathrm{eq}}

\begin{document}
\maketitle

\begin{abstract}
The renewed interest in the foundations of quantum statistical mechanics in recent years has led us to study John von Neumann's 1929 article on the quantum ergodic theorem. We have found this almost forgotten article, which until now has been available only in German, to be a treasure chest, and to be much misunderstood. In it, von Neumann studied the long-time behavior of macroscopic quantum systems. While one of the two theorems announced in his title, the one he calls the ``quantum $H$-theorem,'' is actually a much weaker statement than Boltzmann's classical $H$-theorem, the other theorem, which he calls the ``quantum ergodic theorem,'' is a beautiful and very non-trivial result. It expresses a fact we call ``normal typicality'' and can be summarized as follows: For a ``typical'' finite family of commuting macroscopic observables, every initial wave function $\psi_0$ from a micro-canonical energy shell so evolves that for most times $t$ in the long run, the joint probability distribution of these observables obtained from $\psi_t$ is close to their micro-canonical distribution. 

\medskip

\end{abstract}

\section{Introduction}

In recent years, there has been renewed interest in the foundations of quantum statistical mechanics, see, e.g., \cite{GMM04,PSW06, R08,RDO08,LPSW08}. Our own research in this direction has led us to questions which we later discovered had already been addressed, and some in fact solved, by John von Neumann in his 1929 article on the quantum ergodic theorem (QET) \cite{vN29}. This article concerns the long-time behavior of macroscopic quantum systems, and in particular the approach to thermal equilibrium. We have found the article very useful, and think that it will also be of interest to a wider audience interested in the foundations of quantum statistical mechanics. Here we present an English translation of the 1929 QET article by R.~Tumulka, together with some commentary. In this commentary, we describe von Neumann's results in a non-technical (at least, less technical) way, elaborate on the aspects that we think need elucidation, and put the result into perspective by comparing it to current work on this topic. N.B. All results to date are still far from solving the mathematical problems concerning the quantitative approach to thermal equilibrium of realistic classical or quantum systems. Even less is known rigorously about properties of non-equilibrium systems, e.g., we are not able to derive the heat equation from either classical or quantum mechanics.

Von Neumann's book on the ``Mathematical Foundations of Quantum Mechanics'' \cite{vN32}, published in German in 1932 and in English in 1955, also contains some thermodynamic considerations in Chapter V. This chapter, however, has only little overlap with the QET article, whose content is mentioned only in two brief sentences. ``The reader who is interested in this problem,'' von Neumann writes on page 416 of his book, ``can refer to the treatments in the references'' [i.e., to the QET article]. We actually found the QET article more illuminating than Chapter V of his book.

The QET article is topical also in the following way. There is no consensus about the definition of thermal equilibrium for a quantum (or even a classical) system in microscopic terms; the main divide in the literature lies between a view that can be called the \emph{ensemblist view}, according to which a system is in thermal equilibrium if it is in a mixed state (represented by an ensemble) that is close to the canonical (or micro-canonical) mixed state, and a view that can be called the \emph{individualist view}, according to which a system in a pure state (or a point in phase space) can very well be in thermal equilibrium, depending on the state. The ensemblist view has traditionally prevailed, but the individualist view has gained ground recently (see, e.g., \cite{Deu91,Sre94,T98,GLTZ06,PSW06, R08,RDO08,LPSW08,GLMTZ09b}). While von Neumann's ideas contain elements of both views, the QET is based mainly on the individualist view; indeed, he considered an isolated quantum system described by a pure state that evolves unitarily. We will elaborate on these two views in Section~\ref{sec:mainapproach} below.

The QET article contains two theorems, mentioned already in its title: one von Neumann called the quantum ergodic theorem, the other the quantum $H$-theorem (in analogy to Boltzmann's $H$-theorem in classical mechanics \cite{B96}). These two theorems are so closely related to each other in substance that one and the same proof establishes both of them. For this reason, and because the ``quantum $H$-theorem'' actually asserts much less than Boltzmann's $H$-theorem, we will discuss it only in Section~\ref{sec:mainHtheorem} below and focus otherwise on the QET.

We will convey the content of the QET in Section~\ref{sec:mainqualitative}. It expresses a precise version of a phenomenon we call ``normal typicality'' \cite{GLMTZ09a}: under conditions that are ``typically'' satisfied, every wave function $\psi$ from a micro-canonical energy shell displays the same ``normal'' long-time behavior, viz., for most times $t$ the ``macroscopic appearance'' of $\psi_t$ is the same as that of the micro-canonical ensemble. Here, the macroscopic appearance is expressed in terms of von Neumann's concept of \emph{macroscopic observables}, which was developed for the first time (as far as we are aware) in the QET article and which we will outline in Section~\ref{sec:macroobs}.

The QET provides a condition under which ``normal'' long-time behavior occurs, and it also says that this condition is satisfied for \emph{most} finite families of commuting macroscopic observables (or, in fact, for most Hamiltonians \cite{GLMTZ09a}). It is thus perhaps the first \emph{typicality theorem} in quantum mechanics. Typicality theorems, i.e., statements about most wave functions or most observables or most Hamiltonians, are now widely used. They were crucial to Wigner's work on random Hamiltonians in nuclear physics in the 1950s \cite{Wigner} and are currently used in a great variety of ``random'' systems. Typicality has also been used in recent years in the context of \emph{canonical typicality} (i.e., the fact that, for most wave functions from a narrow energy shell of a large system, the reduced density matrix of a small subsystem is approximately canonical) \cite{GMM04,GLTZ06,PSW06, R07}. For other uses of typicality see, e.g., \cite{GAP,GAPtyp,R08b, SDGP07,BG09,GOZ09}. 

When the QET article was published in 1929, Schr\"odinger wrote an enthusiastic letter to von Neumann \cite{Sch29}. Among other things, he wrote:
\begin{quotation}
Your statistical paper has been of extraordinary interest to me, I am very happy about it, and I'm particularly happy about the gorgeous clarity and sharpness of the concepts and about the careful bookkeeping of what has been achieved at every point.\footnote{Translated from the German by R.~Tumulka.}
\end{quotation}
Schr\"odinger had previously published work \cite{Schr27} on thermodynamic properties of macroscopic quantum systems that one would nowadays regard as a precursor of canonical typicality. A few years later, Pauli and Fierz \cite{PF37} published an alternative proof of the QET which, however, yields weaker error bounds than von Neumann's proof. During the 1930s, the QET was also mentioned in expositions of the foundations of quantum statistical mechanics by Kemble \cite{Kem39} and Tolman \cite[p.~472]{Tol38} (who misattributed it, though, to Pauli and Fierz).

In the 1950s, two articles appeared expressing sharp criticisms of the QET, one by Farquhar and Landsberg \cite{FL57} and one by Bocchieri and Loinger \cite{BL58}. They claimed to have ``mathematically proved the inadequacy of von Neumann's approach'' \cite{BL58} and that ``the von Neumann approach is unsatisfactory'' \cite{FL57}. The authors repeated their criticisms in later publications \cite{BL59,Fbook,Lan05}, calling the QET ``essentially wrong'' \cite{Lan05}, ``seriously flawed'' \cite{Lan05}, and ``devoid of dynamical content'' \cite[p.~166]{Fbook}. However, in these works the QET was mixed up with other statements that indeed are devoid of dynamical content, and the criticisms do not apply to the original QET. Unfortunately, this misunderstanding was not pointed out until recently \cite{GLMTZ09a}; in the 1950s and 1960s, the negative assessment of Farquhar, Landsberg, Bocchieri, and Loinger was widely cited and trusted (e.g., \cite{Lud58a,Lud58b,Lud62,vH59, F60,PS60,J63,Pech84}). In 1962 Ludwig expressed the widespread view in this way \cite{Lud62}:
\begin{quotation}
A short time after the development of quantum mechanics, J.~von Neumann has given a proof of some kind of ergodic theorem. [...] After there was shown by papers of Landsberg and Farquhar and then definitively by a very clear paper of Bocchieri and Loinger that this proof is a physically meaningless one, it is superfluous to go deeper into this proof.
\end{quotation}
As a consequence, the QET was undeservingly forgotten. We elaborate on the
nature of the misunderstanding in \cite{GLMTZ09a} and in
Section~\ref{sec:formost} below. We note, however, that in the 1966 review
by Bocchieri and Prosperi \cite{BP66} of the development of ergodic theory in
quantum mechanics, no criticism of von Neumann based on this
misunderstanding is made.

The remainder of this paper is organized as follows. In Section~\ref{sec:mainqualitative}, we give a qualitative summary of the QET. 
In Section~\ref{sec:classical}, we compare the QET with the situation in classical mechanics and the concept of ergodicity. 
In Section~\ref{sec:formost}, we describe the nature of the widespread misunderstanding of the QET from the 1950s onwards. 
In Section~\ref{sec:mainapproach} we review different definitions of thermal equilibrium and compare the QET to recent works on the approach to thermal equilibrium. 
In Section~\ref{sec:typicality}, we discuss the general relevance of typicality theorems. 
In Section~\ref{sec:mainHtheorem}, we discuss the contents and significance of von Neumann's ``quantum $H$-theorem.'' 
Because the statement of the QET in the QET article is distributed over several places, we formulate it in Appendix~\ref{sec:thm} as a concise and precise mathematical theorem. 
In Appendix~\ref{sec:notation}, we provide a table with von Neumann's notation and elucidate some of his terminology.

\section{Qualitative Summary of the Quantum Ergodic Theorem}
\label{sec:mainqualitative}

\subsection{Setting}
\label{sec:setting}

Von Neumann considered a macroscopic quantum system, confined to a finite volume of space. For the sake of concreteness, we suggest that readers think of a system of $N$ interacting particles, where $N$ is very large (usually larger than $10^{20}$), in a box $\Lambda \subset \RRR^3$. The wave function $\psi_t=\psi(q_1,\ldots,q_N,t)$ evolves according to the Schr\"odinger equation
\be\label{Schr}
i\hbar \frac{\partial\psi}{\partial t} = H\psi_t
\ee
with $H$ the Hamiltonian of the system. (Von Neumann used the opposite sign in the Schr\"odinger equation, writing $-i$ instead of $i$. The form \eqref{Schr} is nowadays standard.) As usual, $\psi_0$ (and thus $\psi_t$) should be normalized,
\be
\|\psi_0\|^2=\int_\Lambda \cdots \int_\Lambda |\psi_0|^2 \, d^3q_1 \cdots d^3q_N =1\,. 
\ee
It follows from the confinement to a finite volume that $H$ has discrete energy levels, which we denote $E_\alpha$ (see Appendix~\ref{sec:notation} for a list of von Neumann's notation). Let $\{\phi_\alpha\}$ be an orthonormal basis of the system's total Hilbert space $\Hilbert_{\total}$ consisting of eigenfunctions of $H$,
\be
H\phi_\alpha=E_\alpha\phi_\alpha\,.
\ee
Considering only Hamiltonians that are bounded from below, there will be only finitely many eigenvalues (with multiplicity) below any given value, so we can order them so that $E_0\leq E_1 \leq E_2 \leq \ldots$.

Von Neumann considered further (what amounts to) a partition of the energy axis, or rather of the relevant half-axis $[E_0,\infty)$, into disjoint intervals $\mathscr{I}_a=[\mathscr{E}_a,\mathscr{E}_{a+1})$ (with $\mathscr{E}_0=E_0$ and $\mathscr{E}_a<\mathscr{E}_{a+1}$) that are large on the microscopic scale (so that each contains many eigenvalues $E_\alpha$), but small on the macroscopic scale (so that different energies in one interval are not macroscopically different).\footnote{In von Neumann's words (Section 1.2): ``With a certain (reduced) accuracy, [it is] possible [to measure energy with macroscopic means], so that the energy eigenvalues [...] can be collected in groups [...] in such a way that all [eigenvalues in the same group] are close to each other and only those [in different groups] can be macroscopically distinguished.''} Such an interval is called a \emph{micro-canonical energy shell}, an expression that is also often used to refer to the subspace $\Hilbert_{\mathscr{I}_a}\subseteq \Hilbert_{\total}$ spanned by the $\phi_\alpha$ with $E_\alpha \in \mathscr{I}_a$.

\subsection{Macroscopic Observables}
\label{sec:macroobs}

We now turn to the mathematical structure that encodes the concept of ``macroscopic'' in von Neumann's article: 
a decomposition of the Hilbert space $\Hilbert$ into mutually orthogonal subspaces $\Hilbert_\nu$, 
\be\label{orthodecomp} 
\Hilbert = \bigoplus_\nu \Hilbert_\nu\,, 
\ee 
such that each $\Hilbert_\nu$ corresponds to a different macro-state $\nu$. We call the $\Hilbert_\nu$ the ``macro-spaces'' and write $\decomp$ for the family $\{\Hilbert_\nu\}$ of subspaces, called a ``macro-observer'' in von Neumann's paper, and $P_\nu$ for the projection to $\Hilbert_\nu$. We use the notation
\be
\dd_\nu =\dim \Hilbert_\nu\,.
\ee

For the sake of simplicity, we focus on only one micro-canonical interval $\mathscr{I}_a=[\mathscr{E}_a,\mathscr{E}_{a+1})$ and simply 
consider $\Hilbert_{\mathscr{I}_a}$, rather than $\Hilbert_{\total}$, as our Hilbert space $\Hilbert$. In particular, we take \eqref{orthodecomp} to be a decomposition of the energy shell $\Hilbert_{\mathscr{I}_a}$; this decomposition can be regarded as analogous to a partition of the energy shell in a classical phase space. 
Let $\D=\dim\Hilbert$, i.e., the number of energy levels, including multiplicities, between $\mathscr{E}_a$ and $\mathscr{E}_{a+1}$. This number is finite but huge---usually greater than $10^{10^{20}}$ when the number $N$ of particles is greater than $10^{20}$. 

The \emph{micro-canonical density matrix} $\rho_{\mc}$ is the projection to $\Hilbert$ times a normalization factor $1/\D$,
\be
\rho_{\mc} = \frac{1}{D} \sum_{\alpha:\phi_\alpha\in\Hilbert} \pr{\phi_\alpha}\,,
\ee
and the \emph{micro-canonical average} of an observable $A$ on $\Hilbert$ is given by
\be\label{mcav}
\tr(\rho_{\mc}A)=\frac{\tr A}{\D}\,.
\ee
This value can also be obtained as the average of the values $\scp{\phi_\alpha}{A|\phi_\alpha}$, with equal weights $1/\D$, over those $\alpha$ with $E_\alpha\in[\mathscr{E}_a,\mathscr{E}_{a+1})$. Alternatively, it can also be obtained as the average of the values $\scp{\varphi}{A|\varphi}$ with $\varphi$ uniformly distributed over the unit sphere
\be
\sphere(\Hilbert) = \{\varphi\in\Hilbert:\|\varphi\|=1\}\,.
\ee
For this uniform distribution, the probability that $\varphi\in B\subseteq \sphere(\Hilbert)$ is the $(2\D-1)$-dimensional surface area of $B$ times a normalization factor.

Von Neumann motivated the decomposition \eqref{orthodecomp} by beginning with a
family of operators corresponding to coarse-grained macroscopic observables
and arguing that by ``rounding'' the operators, the family can be converted
to a family of operators $M_1,\ldots,M_\ell$ that commute with each other,
have pure point spectrum, and have huge degrees of degeneracy. 
A macro-state can then be characterized by a list $\nu=(m_1,\ldots,m_\ell)$ of
eigenvalues $m_i$ of the $M_i$, and corresponds to the subspace
$\Hilbert_\nu \subseteq \Hilbert_{\total}$ containing the simultaneous eigenvectors
of the $M_i$ with eigenvalues $m_i$; that is, $\Hilbert_\nu$ is the
intersection of the respective eigenspaces of the $M_i$ and $\dd_\nu$ is
the degree of simultaneous degeneracy of the eigenvalues
$m_1,\ldots,m_\ell$.\footnote{For a notion of macro-spaces that does not require that
  the corresponding macro-observables commute, see \cite{DRMN06}, in
  particular Section 2.1.1.} 
If any of the $\Hilbert_\nu$ has dimension 0, i.e., if a particular combination of eigenvalues of the $M_i$ does not occur, then we delete it from the family $\decomp$. 

As an example of the ``rounding'' of macroscopic observables, von Neumann points out that when we simultaneously measure the position and momentum of a macroscopic body, the experiment corresponds not to the exact center-of-mass position and total momentum observables but to two commuting observables approximating these.\footnote{In von Neumann's words (Section 0.2): ``In a macroscopic measurement of coordinate and momentum (or two other quantities that cannot be measured simultaneously according to quantum mechanics), really two physical quantities are measured simultaneously and exactly, which however are not exactly coordinate and momentum.''}
Correspondingly, the distance between neighboring eigenvalues of $M_i$ represents the inaccuracy of the measurement. As further examples of macro-observables, we may consider similar approximations to the number of particles in the left half of the box $\Lambda$ divided by the total number of particles, or to the $z$-component of the magnetization (i.e., the total magnetic $z$-moment $\sum_{i=1}^N \sigma_{z,i}$, where $\sigma_{z,i}$ is the third Pauli matrix acting on the $i$-th particle).

Von Neumann's reasoning, that macroscopic observables can be taken to commute, has inspired research about whether for given operators $A_1,\ldots,A_\ell$ whose commutators are small one can find approximations $M_i\approx A_i$ that commute exactly; the answer is, for $\ell\geq 3$ and general $A_1,\ldots,A_\ell$, no \cite{Choi88}. Also for two operators, the question is mathematically non-trivial; for recent results see \cite{Lin95,Has09}, for an overview see \cite{HL09}. Von Neumann gave an example of such an approximation in his Section 0.2, starting from the position and momentum operators $\Qq_k$ and $\Pp_k$ of quantum mechanics and providing new operators $\Qq_k'$ and $\Pp_k'$ that commute; however, it is perhaps not a good example because, somewhat contrary to what he suggested in his Section 0.2, it is not clear why $\Qq_k-\Qq_k'$ or $\Pp_k-\Pp_k'$ should have small operator norm (or which other norm could be relevant) [E.~Carlen and M.~Hastings, personal communication].\footnote{Concerning his construction of $\Qq_k'$ and $\Pp_k'$, we also note that the constant $C$ (that controls the spread of the joint eigenfunctions $\varphi_n$ of $\Qq_k'$ and $\Pp_k'$ in the position and momentum representations) was specified by von Neumann in Footnote 9 of his 1929 QET article as ``$C<3.6$'' but in his 1932 book \cite{vN32} as ``$C\sim 60$,'' so maybe the bound $C<3.6$ was incorrectly calculated. Bourgain \cite{Bou88} has improved the constant to $C=1$; i.e., he has found another basis $\tilde{\varphi}_n$ that achieves with $C=1$ (the bound suggested by the uncertainty principle) what von Neumann asserted of his choice of basis $\varphi_n$ described in his Footnote 10.}

When von Neumann considered the commuting macro-observables $M_1,\ldots,M_\ell$, he had in mind that one of them, 
say $M_1$, is the ``macroscopic energy,'' which can be thought of as obtained from $H$ by coarse-graining in agreement with the partition of the energy axis into the micro-canonical intervals $\mathscr{I}_a$,
\be\label{M1}
M_1 = \sum_{\alpha} f_1(E_\alpha) \, \pr{\phi_\alpha}
\ee
with $f_1$ the appropriate step function given by
\be
f_1(E)=\frac{\mathscr{E}_a + \mathscr{E}_{a+1}}{2}
\quad \text{for} \quad E\in\mathscr{I}_a=[\mathscr{E}_a,\mathscr{E}_{a+1})\,.
\ee
Since the $M_i$ commute with one another, every $M_i$ commutes with the \emph{coarse-grained} energy $M_1$, but generally not with $H$, so it is generally not a conserved quantity.\footnote{In von Neumann's words (Section 1.2): ``In general, [...] $H$ is not a linear combination of the [projections to the joint eigenspaces of all $M_i$], since the energy is not a macroscopic quantity, as it cannot be measured with absolute precision with macroscopic means'' [as the $M_i$ can].} 

Since all $M_i$ commute, every $M_i$ maps the energy shell $\Hilbert$ to itself, and we can (and will) regard the $M_i$ as operators on $\Hilbert$. Thus, our assumption is satisfied
that each of the macro-spaces $\Hilbert_\nu$ either lies in $\Hilbert$ or is orthogonal to $\Hilbert$. The size of $\dd_\nu$ is in practice also of the rough order $10^{10^{20}}$, though often very much smaller than $\D$. (Note that, e.g., $10^{0.9999 \times 10^{20}}$ is smaller than $10^{10^{20}}$ by a factor of $10^{10^{16}}$).

\subsection{Statement of the Quantum Ergodic Theorem}
\label{sec:qualitative}

We now have the ingredients---$H$, $\Hilbert$, and $\decomp$---to formulate the QET. Despite the name, the property described in the QET is not precisely analogous to the standard notion of \emph{ergodicity} as used in classical mechanics and the mathematical theory of dynamical systems. That is why we prefer to call quantum systems with the relevant property ``normal'' rather than ``ergodic.''\footnote{This terminology is inspired by the concept of a \emph{normal real number}, which is ``a real number whose digits in every base show a uniform distribution, with all digits being equally likely, all pairs of digits equally likely, all triplets of digits equally likely, etc.. While a general proof can be given that almost all numbers are normal, this proof is not constructive [...]. 
It is for instance widely believed that the numbers $\sqrt{2}$, $\pi$, and $e$ are normal, but a proof remains elusive.'' \cite{normalnumber}} Let us proceed towards a description of this property.

Any wave function $\psi\in\Hilbert$ with $\|\psi\|=1$ defines a probability distribution over all macro-states $\nu$; namely, the probability associated with $\nu$ is
\be\label{probnupsi}
\|P_\nu \psi\|^2 = \scp{\psi}{P_\nu|\psi}\,.
\ee
(Recall that $P_\nu$ is the projection to $\Hilbert_\nu$.)
This is the probability of obtaining, in a joint measurement of the macro-observables $M_1,\ldots,M_\ell$ on a system in state $\psi$, the outcomes $(m_1,\ldots,m_\ell)$ corresponding to $\nu$. Similarly, the micro-canonical density matrix $\rho_{\mc}$ defines a probability distribution over all macro-states $\nu$; namely, the probability associated with $\nu$ is
\be\label{probnumc}
\tr(\rho_{\mc} P_\nu) = \frac{\dd_\nu}{\D}\,.
\ee

\begin{claim}
For most wave functions $\psi$ from the unit sphere in the micro-canonical subspace $\Hilbert$, the distribution \eqref{probnupsi} associated with $\psi$ is close to the micro-canonical distribution \eqref{probnumc}.
\end{claim}

The reference to ``most'' is intended to convey that the subset of the unit sphere in $\Hilbert$ containing those $\psi$ for which \eqref{probnupsi} is close (in some precise sense) to \eqref{probnumc} has measure arbitrarily close to 1, provided each of the $\dd_\nu$ is sufficiently large. Here, the ``measure'' corresponds to the uniform distribution over the unit sphere. Claim 1 follows from the fact, proven by von Neumann in his appendices A.1--A.3, that if $\Hilbert_\nu$ is any fixed subspace of $\Hilbert$ of dimension $\dd_\nu$ and $\varphi$ is a random vector with uniform distribution on the unit sphere in $\Hilbert$ then
\be\label{EEE}
\EEE \|P_\nu \varphi\|^2 = \frac{\dd_\nu}{\D}\,,\quad
\mathrm{Var} \|P_\nu\varphi\|^2 = \EEE\Bigl(\|P_\nu\varphi\|^2-\frac{\dd_\nu}{\D}\Bigr)^2
<\frac1{\dd_\nu}\Bigl(\frac{\dd_\nu}{\D}\Bigr)^2\,.
\ee
Here, $\EEE$ denotes the expected value and $\mathrm{Var}$ the variance of a random variable. Thus, the first equation in \eqref{EEE} says that the value \eqref{probnupsi} associated with $\psi$, when averaged over the unit sphere, yields the micro-canonical value \eqref{probnumc}, and the second equation says that the standard deviation of the random variable $\|P_\nu \varphi\|^2$ is small, in fact much smaller than its average, provided $\dd_\nu \gg 1$. It then follows from Chebyshev's inequality that the probability that $\|P_\nu\varphi\|^2$ deviates much from its expectation value $\dd_\nu/\D$ is small. That is, in the language of measure theory, the set of $\psi$s for which $\|P_\nu\psi\|^2$ deviates much from the micro-canonical value is small, which was what was claimed.

As a consequence of Claim 1, most wave functions $\psi$ are such that for each of the macroscopic observables $M_1,\ldots,M_\ell$---and, in fact, for every function $f(M_1,\ldots,M_\ell)$, i.e., for every element of the algebra generated by $M_1,\ldots,M_\ell$---the probability distribution that $\psi$ defines on the spectrum of the observable is close to the one defined by the micro-canonical density matrix. Put loosely, most pure states in $\Hilbert$, when looked at macroscopically, look like the micro-canonical mixed state. It is clear that Claim 1 cannot be true for \emph{all} (rather than most) wave functions, as one can easily provide examples of wave functions whose distribution is not close to the micro-canonical one: say, $\psi\in\Hilbert_\nu$ for one particular $\nu$. 

Let us consider now the time evolution of some initial $\psi$ and ask whether
\be\label{approx}
\|P_\nu \psi_t\|^2 \approx \frac{\dd_\nu}{\D}
\quad \text{for all }\nu
\ee
will hold for most times $t$.\footnote{When saying ``most $t$,'' we have in mind most $t>0$, but the QET and our other statements are equally true for most $t<0$, as long as the system was and remains isolated.} This may seem like a plausible behavior in view of Claim 1. In fact, from Claim 1 it follows rather easily that \eqref{approx} holds for \emph{most} initial wave functions $\psi_0$ and \emph{most} times $t$. The QET goes further. It asserts that, for certain systems,  \eqref{approx} holds for \emph{all} initial wave functions $\psi_0$ for most times $t$. This is important because one may expect \emph{most} wave functions to represent microscopic states of thermal equilibrium, while states of non-equilibrium should form a very small minority. Thus, if we are interested in the evolution towards equilibrium, we are specifically interested in the question whether non-equilibrium states will evolve towards equilibrium, and hence we cannot be satisfied with statements about \emph{most} wave functions because such statements need not apply to the non-equilibrium wave functions.

Let us put this differently. We call a system, defined by $H$, $\Hilbert$, $\decomp$, and $\psi_0\in\Hilbert$, \emph{normal} if and only if \eqref{approx} holds for most $t$. The QET provides conditions under which a system is normal for every initial state vector $\psi_0$. Furthermore, the QET asserts \emph{normal typicality}, i.e., that typical macroscopic systems are normal for every $\psi_0$; more precisely, that for \emph{most} choices of $\decomp$, macroscopic systems are normal for every $\psi_0$. The result is, in fact, equivalent to the statement that for most Hamiltonians, macroscopic systems are normal for every $\psi_0$ \cite{GLMTZ09a}.\footnote{The concept of ``most $\decomp$'' refers to the uniform distribution over all orthogonal decompositions \eqref{orthodecomp} such that $\dim \Hilbert_\nu=\dd_\nu$. When talking about ``most Hamiltonians'' we refer to the uniform distribution over all Hamiltonians with given eigenvalues. Both distributions are marginals of images of the Haar measure over the group of unitary $\D\times\D$ matrices; for their full definitions see \cite{GLMTZ09a} or the QET article.}
It thus provides reasons to believe (but no proof) that macroscopic systems in practice are normal.
 
Before we give an informal statement of the QET, we formulate two conditions involved in it. First, following von Neumann, we say that a Hamiltonian $H$ with   
eigenvalues $E_1,\ldots,E_\D$ has no resonances if and only if
\be\label{noresonance}
E_{\alpha}-E_{\beta} \neq E_{\alpha'}-E_{\beta'}
\text{ unless }\begin{cases}\text{either } \alpha= \alpha' \text 
{ and } \beta= \beta' \\
\text{or }\alpha=\beta \text{ and }\alpha'=\beta'\,.\end{cases}
\ee
In words, this means that the energy differences are non-degenerate. It implies in particular that the energy levels themselves are non-degenerate, but is a stronger condition. The other condition is a technical one and can be stated as follows. For a given $H$ and $\decomp=\{\Hilbert_\nu\}$, let
\be\label{cond2c}
F_\nu(H,\decomp) = 
\max_{\alpha\neq \beta}
\bigl|\scp{\phi_\alpha}{P_\nu|\phi_\beta}\bigr|^2 
+ \max_\alpha 
\Bigl(\scp{\phi_\alpha}{P_\nu|\phi_\alpha}-\frac{\dd_\nu}{\D}\Bigr)^2\,.
\ee
The condition, to which we will simply refer as ``condition \eqref{cond2a},'' is that
\be\label{cond2a}
F_\nu(H,\decomp) 
\text{ is sufficiently small for every }\nu\,.
\ee

\bigskip

\noindent\textbf{Informal statement of the QET.} {\it (For a fully precise statement  
see Appendix~\ref{sec:thm} below.) Let $\Hilbert$ be any Hilbert space of finite dimension $\D$, let $\decomp=\{\Hilbert_\nu\}$ be an orthogonal decomposition of $\Hilbert$ with $\dim\Hilbert_\nu=\dd_\nu$, and let the Hamiltonian $H$ be a self-adjoint operator on $\Hilbert$ without resonances. If $H$ and $\decomp$ satisfy condition \eqref{cond2a} then, for every wave function $\psi_0\in\Hilbert$ with $\|\psi\|=1$, the system is normal, i.e., \eqref{approx} holds most of the  
time. Moreover, for sufficiently large $\dd_\nu$s with $\sum_\nu \dd_\nu=\D$, most families $\decomp = \{\Hilbert_\nu\}$ of mutually orthogonal subspaces $\Hilbert_\nu$ with $\dim\Hilbert_\nu= \dd_\nu$ are such that condition \eqref{cond2a} is satisfied (and thus the system is normal for every $\psi_0$).}

\bigskip

(It is easy to understand the role of the second term on the right hand side of \eqref{cond2c}: When we want to ensure that the system is normal for \emph{every} wave function, then we need to ensure that it is for energy eigenfunctions $\phi_\alpha$. Since the time evolution of $\phi_\alpha$ is stationary, it can only be normal if $\scp{\phi_\alpha}{P_\nu|\phi_\alpha}\approx \dd_\nu/\D$, which corresponds to the smallness of the second term in \eqref{cond2c}.)

Here is another way of expressing the QET. 
Let us denote the long-time average of a function $f(t)$ by an overbar,
\be
\overline{f(t)} = \lim_{T\to\infty} \frac{1}{T} \int_0^T dt \, f(t)\,.
\ee
(All statements remain valid if we include negative times and set
\be
\overline{f(t)} = \lim_{T\to\infty} \frac{1}{2T} \int_{-T}^T dt \, f(t)\,.) 
\ee
Then a system is normal if, for every $\nu$, the time average $\overline{\|P_\nu\psi_t\|^2}$ is close to $\dd_\nu/\D$ \emph{and} the time variance of $\|P_\nu\psi_t\|^2$ is small; equivalently, a system is normal if, for every $\nu$, the expression
\be\label{expr1}
\bigl(\|P_\nu\psi_t\|^2-\dd_\nu/\D\bigr)^2
\ee
has small time average. The QET asserts that the time average of \eqref{expr1} is no greater than \eqref{cond2c} (independently of $\psi_0$), and, moreover, that the average of \eqref{cond2c} over all $\decomp$ with $\dim\Hilbert_\nu=\dd_\nu$ is small when $\dd_\nu$ is sufficiently large.

More detailed discussions of the QET have been provided by Pauli and Fierz in 1937 \cite{PF37} and by Jancel in 1963 \cite{J63}; see also \cite{GLMTZ09a}.

\section{Comparison With Classical Mechanics}
\label{sec:classical}

For a classical Hamiltonian system, we denote a point in phase space by
\be
X=(\vq_1,\ldots,\vq_N,\vp_1,\ldots,\vp_N)\,.
\ee
The time evolution of the micro-state $X$ is given by the solution of the Hamiltonian equations of motion, which sends $X$ (at time $0$) to $X_t$ (at time $t$), $t\in\RRR$. This dynamics preserves the Liouville phase-space volume.

Instead of the orthogonal decomposition of $\Hilbert$ into subspaces $\Hilbert_\nu$ we consider a partition of an energy shell $\Gamma$ in phase space $\RRR^{6N}$,
\be\label{Gammadef}
\Gamma = \{X:\mathscr{E}_a \leq H(X)< \mathscr{E}_{a+1}\}\,,
\ee
into regions $\Gamma_\nu$ corresponding to different macro-states $\nu$,
\be\label{Gammanu}
\Gamma = \bigcup_\nu \Gamma_\nu\,,
\ee
i.e., if the micro-state $X$ of the system is in $\Gamma_\nu$ then the macro-state of the system is $\nu$. Let $\mu_{\mc}$ denote the micro-canonical distribution, i.e., the uniform distribution (= normalized Liouville phase space volume) on $\Gamma$. Then with each macro-state $\nu$ there is associated the micro-canonical probability $\mu_{\mc}(\Gamma_\nu)$.

A crucial difference between a quantum and a classical system is that $\psi$ can be a superposition of contributions from several $\Hilbert_\nu$s whereas $X$ always lies in one and only one of the $\Gamma_\nu$. As a consequence, a single phase point $X$ does not provide a nontrivial probability distribution over the $\nu$s, and there is no statement analogous to \eqref{approx} in classical mechanics. One can only ask about the fraction of time that $X_t$ spends in various $\Gamma_\nu$s, and to this question we turn in the following subsection.

\subsection{Ergodicity}
\label{sec:ergodic}

As we mentioned already, normality---the property relevant to the QET---is not really analogous to ergodicity. Nevertheless, to formulate a quantum analog of ergodicity was von Neumann's motivation for the QET. 

Let us recall the concept of ergodicity (called ``quasi-ergodicity'' in the 1920s) in statistical mechanics. Let $\Gamma_{\mathscr{E}}$ denote the energy surface,
\be\label{GammaEdef}
\Gamma_{\mathscr{E}}=\{X\in\RRR^{6N}:H(X)=\mathscr{E}\}\,,
\ee
and $\mu_{\mathscr{E}}$ the (micro-canonical) invariant measure on $\Gamma_{\mathscr{E}}$ defined to be the limit of the normalized phase space volume measure $\mu_{\mc}$ as both $\mathscr{E}_a\to \mathscr{E}$ and $\mathscr{E}_{a+1}\to\mathscr{E}$; in fact, $\mu_{\mathscr{E}}$ is the surface area measure re-weighted with the inverse norm of the gradient of the Hamiltonian function and normalized. The dynamics generated by $H$ on $\Gamma_{\mathscr{E}}$ is \emph{ergodic} if it has no non-trivial (measurable) constants of the motion. As a consequence of Birkhoff's ergodic theorem \cite{Bir31}, this is equivalent to the following: the fraction of time that the phase point $X_t$ spends in a region $B\subseteq \Gamma_{\mathscr{E}}$ is in the long run proportional to the size of the region, $\mu_{\mathscr{E}}(B)$, for almost every $X_0\in\Gamma_{\mathscr{E}}$. (``Almost every'' means that the set of exceptions has measure zero; this is different from ``most,'' which conveys that the set of exceptions has small measure---but usually not zero.) Equivalently, time averages coincide with phase-space averages (over the energy surface). Let $\delta_{X_t}$ denote the delta measure concentrated at the phase point $X_t$. Then ergodicity is equivalent to
\be\label{ergodic}
\overline{\delta_{X_t}} =\mu_{\mathscr{E}}
\ee
(with the time average understood in the sense of weak convergence) for almost every $X_0\in\Gamma_{\mathscr{E}}$.

In quantum mechanics, if we regard a pure state $|\psi_t\rangle \langle\psi_t|$ as analogous to the pure state $\delta_{X_t}$ and $\rho_{\mc}$ as analogous to $\mu_{\mathscr{E}}$, the statement analogous to \eqref{ergodic} reads
\be\label{Qergodic}
\overline{\ket{\psi_t}\bra{\psi_t}} = \rho_{\mc}\,.
\ee
As pointed out by von Neumann in his QET article, the left hand side always exists\footnote{This existence statement also follows, at least for \emph{almost} every $\psi_0$, from the (classical) ergodic theorems of Birkhoff \cite{Bir31} and von Neumann \cite{vN32a}; however, the QET article appeared two years earlier.} and can be computed as follows. If $\psi_0$ has coefficients $c_\alpha=\scp{\phi_\alpha}{\psi_0}$ in the energy eigenbasis $\{\phi_\alpha\}$,
\be
\psi_0 = \sum_{\alpha=1}^\D c_\alpha \ket{\phi_\alpha}\,,
\ee
then
\be
\psi_t = \sum_{\alpha=1}^\D e^{-iE_\alpha t/\hbar} c_\alpha \ket{\phi_\alpha}\,,
\ee
and thus
\be
\overline{\ket{\psi_t}\bra{\psi_t}}=
\sum_{\alpha,\beta} \overline{e^{-i(E_\alpha-E_\beta)t/\hbar}} c_\alpha c_\beta^* \ket{\phi_\alpha}\bra{\phi_\beta}\,.
\ee
Suppose that $H$ is non-degenerate; then $E_\alpha-E_\beta$ vanishes only for $\alpha=\beta$, so the time averaged exponential is $\delta_{\alpha\beta}$, and we have that
\be
\overline{\ket{\psi_t}\bra{\psi_t}}
=\sum_{\alpha} |c_\alpha|^2 \ket{\phi_\alpha}\bra{\phi_\alpha}\,.
\ee
Thus, the case \eqref{Qergodic} occurs only for those special wave functions that have $|c_\alpha|^2=1/\D$ for all $\alpha$. That is, the property of a quantum system that is the most obvious analog of ergodicity is almost never satisfied.

One can draw other analogies, though, by focusing just on the macroscopic appearance, understood in terms of the macroscopic observables $M_1,\ldots,M_\ell$ mentioned in Section~\ref{sec:setting} above and the orthogonal decomposition $\decomp = \{\Hilbert_\nu\}$ they define. We say that two density matrices $\rho$ and $\rho'$ are \emph{macroscopically equivalent}, in symbols
\be
\rho \stackrel{\decomp}{\sim} \rho'\,,
\ee
if and only if
\be
\tr(\rho P_\nu) \approx \tr(\rho' P_\nu)
\ee
for all $\nu$. For example, $\ket{\psi}\bra{\psi}\stackrel{\decomp}{\sim}\rho_{\mc}$ if and only if
\be\label{approx1}
\|P_\nu \psi\|^2 \approx \frac{\dd_\nu}{\D}
\ee
for all $\nu$. This is exactly the condition considered in Claim 1 in Section~\ref{sec:qualitative}, so this is true of most $\psi$ (provided $\dd_\nu \gg 1$). 
Returning to the time average, we obtain that $\overline{\ket{\psi_t}\bra{\psi_t}}\stackrel{\decomp}{\sim}\rho_{\mc}$ if and only if
\be\label{mcappear}
\sum_\alpha |c_\alpha|^2 \scp{\phi_\alpha}{P_\nu|\phi_\alpha} \approx \frac{\dd_\nu}{\D}
\ee
for all $\nu$. Condition \eqref{mcappear} is satisfied for every $\psi_0\in\sphere(\Hilbert)$  if and only if
\be
\scp{\phi_\alpha}{P_\nu|\phi_\alpha} \approx \frac{\dd_\nu}{\D}
\ee
for every $\alpha$ and $\nu$, a condition on $H$ and $\decomp$
that follows from \eqref{cond2a} and thus is, according to the QET, 
typically obeyed (see also \cite{BP66}). The analogy between
$\overline{\ket{\psi_t}\bra{\psi_t}}\stackrel{\decomp}{\sim}\rho_{\mc}$ and
ergodicity lies in the fact that the time average of a pure state in a
sense agrees with the micro-canonical ensemble (with two differences:
that the agreement is only an approximate agreement on the macroscopic
level, and that it typically holds for \emph{every}, rather than \emph{almost every}, pure state).

However, even more is true for many quantum systems: Not just the time average but even $\ket{\psi_t}\bra{\psi_t}$ itself is macroscopically equivalent to $\rho_{\mc}$ for most times $t$, as expressed in \eqref{approx}. Thus, normality is in part stronger than ergodicity (it involves no time averaging) and in part weaker (it involves only macroscopic equivalence); in short, it is a different notion. In von Neumann's words (first paragraph of his Section 0.5):
\begin{quotation}
[T]he agreement between time and microscopic average should only be required for macroscopic quantities. This weakening comes together with an essential strengthening that is made possible only by using the macroscopic perspective. Namely, we will show that for every state of the system the value of each (macroscopically measurable) quantity not only has time mean equal to the micro-canonical mean, but furthermore has small spread, i.e., the times at which the value deviates considerably from the mean are very infrequent.
\end{quotation}

\subsection{Ergodic Components of the Schr\"odinger Evolution}
\label{sec:ercomp}

Every dynamical system whose dynamics leaves invariant a probability distribution $\mu$ can be partitioned into its ergodic components \cite{Sin76}. That is, its phase space $\Gamma$ can be partitioned in a (more or less) unique way into disjoint subsets, $\Gamma =  \cup_s T_s$ and $T_s\cap T_{s'}=\emptyset$ for $s\neq s'$, so that each $T_s$ is invariant under the dynamics, and the dynamics is ergodic on $T_s$ (equipped with a probability measure $\mu_s$ that it inherits from $\mu$).

In Section 0.4 of the QET article, von Neumann identifies the ergodic components of the Schr\"odinger dynamics, regarded as a dynamical system on the unit sphere of the Hilbert space $\Hilbert_{\total}$, at least when $\Hilbert_{\total}$ is finite dimensional and the eigenvalues of $H$ are linearly independent over the rational numbers (which is the generic case). Here, the invariant distribution is the uniform distribution over the unit sphere, the parameter $s$ is a sequence $(r_\alpha)$ of radii, one for each energy level, and $T_s=T_{r_1,r_2,\ldots}$ is the torus defined by these radii,
\be\label{torus}
T_{r_1,r_2,\ldots}=\Bigl\{\sum_\alpha r_\alpha \, e^{i\theta_\alpha} \ket{\phi_\alpha}: 
0\leq \theta_\alpha<2\pi \Bigr\}\,.
\ee

\section{Misunderstanding in the 1950s}
\label{sec:formost}

As noted before, the QET was widely dismissed after undeserved criticisms in \cite{FL57,BL58} arising from a wrong idea of what the QET asserts. In this section we point out the nature of the misunderstanding. Let $p(\decomp,\psi_0)$ be the statement that the system with initial wave function $\psi_0$ is normal with respect to $\decomp$ (i.e., that \eqref{approx} holds most of the time). The misunderstanding of Bocchieri and Loinger \cite{BL58} consists of replacing the statement
\be\label{mostDallpsip}
\text{for most }\decomp:\: \text{for all }\psi_0:\: p(\decomp,\psi_0)\,,
\ee
which is part of the QET, with the inequivalent (in fact, weaker) statement 
\be\label{allpsimostDp}
\text{for all }\psi_0:\: \text{for most }\decomp:\: p(\decomp,\psi_0)\,.
\ee
To see that these two statements are indeed inequivalent, let us illustrate the difference between ``for most $x$: for all $y$: $p(x,y)$'' and ``for all $y$: for most $x$: $p(x,y)$'' by two statements about a company:
\be\label{Sa}
\mbox{
\begin{minipage}{0.5\textwidth}
\textit{Most employees are never ill.}
\end{minipage}}
\ee
\be\label{Sb}
\mbox{
\begin{minipage}{0.5\textwidth}
\textit{On each day, most employees are not ill.}
\end{minipage}}
\ee
When $x$ ranges over employees, $y$ over days, and $p(x,y)$ is the statement ``Employee $x$ is not ill on day $y$'' then  ``for most $x$: for all $y$: $p(x,y)$'' is \eqref{Sa} and ``for all $y$: for most $x$: $p(x,y)$'' is \eqref{Sb}. It is easy to understand that \eqref{Sa} implies \eqref{Sb}, and \eqref{Sb} does not imply \eqref{Sa}, as there is the (very plausible) possibility that most employees are sometimes ill, but not on the same day. Von Neumann was clearly aware of the difference between \eqref{mostDallpsip} and \eqref{allpsimostDp}, as his footnote 37 in Section 3.1 shows:
\begin{quotation}
Note: what we have shown is not that for every given $\psi$ or $\Aa$ the ergodic theorem and the $H$-theorem hold for most $\omega_{\lambda,\nu,a}$ but that for most $\omega_{\lambda,\nu,a}$ they are universally valid, i.e., for all $\psi$ and $\Aa$. The latter is, of course, much more [i.e., much stronger] than the former.
\end{quotation}
Also Schr\"odinger, by the way, was aware that von Neumann had proven \eqref{mostDallpsip}, as his 1929 letter to von Neumann \cite{Sch29} shows:
\begin{quotation}
You can show: \emph{if this rotation} [i.e., the unitary operator mapping an eigenbasis of $H$ to a joint eigenbasis of $M_1, \ldots, M_\ell$] \emph{is large enough} then the theorem holds with arbitrary accuracy. You can show further: the overwhelming majority of the \emph{conceivable} rotations \emph{is} indeed large enough---where the ``overwhelming majority'' is defined in an appropriate, rotation-invariant way. Given \emph{such} a rotation, then the theorem holds \emph{for every psi}.\footnote{Translated from the German by R. Tumulka.}
\end{quotation}

To see how \eqref{mostDallpsip} and \eqref{allpsimostDp} are connected to the calculations in the QET article, as well as those of Bocchieri and Loinger \cite{BL58}, we note that, as mentioned earlier, the normality of $\psi_0$ with respect to $\decomp$ (i.e., the statement $p(\decomp,\psi_0)$) is equivalent to the statement that, for every $\nu$,
\be\label{expr4}
\overline{\bigl(\|P_\nu\psi_t\|^2 -\dd_\nu/\D\bigr)^2}
\ee
is small. As a straightforward calculation shows (see \cite{GLMTZ09a} or the QET article), the quantity \eqref{expr4} is, for all $\psi_0$, less than or equal to the non-negative quantity $F_\nu(H,\decomp)$ defined in \eqref{cond2c}, which is independent of $\psi_0$. This calculation is von Neumann's argument showing that smallness of $F_\nu=F_\nu(H,\decomp)$ implies normality for every $\psi_0$. The main work involved in proving the QET, though, is to show that $F_\nu$ is small for most $\decomp$, and that is done by showing that the average of $F_\nu$ over all $\decomp$ is small. Bocchieri and Loinger \cite{BL58} considered, instead of the two propositions that
\be\label{prop1}
\eqref{expr4} \leq F_\nu
\ee
and that
\be\label{prop2}
\text{the $\decomp$-average of $F_\nu$ is small,}
\ee
the one proposition that 
\be\label{prop3}
\text{the $\decomp$-average of \eqref{expr4} is small.}
\ee
It can be proven easily that \eqref{prop3} is true for all $\psi_0$, provided the $\dd_\nu$ are sufficiently large, by changing the order of the two operations of taking the time average and taking the $\decomp$-average \cite{BL58,GLMTZ09a}. However, this statement implies only \eqref{allpsimostDp}, and not the stronger statement \eqref{mostDallpsip} needed for the QET. Indeed, (assuming the $\dd_\nu$ are sufficiently large) it follows that for all $\psi_0$ it is true of most $\decomp$ and most $t$ that $\|P_\nu\psi_t\|^2\approx \dd_\nu/\D$; this is \eqref{allpsimostDp}. In contrast, the two propositions \eqref{prop1} and \eqref{prop2} yield that for most $\decomp$ it is true of all $\psi_0$ that, for most $t$, $\|P_\nu\psi_t\|^2 \approx \dd_\nu/\D$; this is \eqref{mostDallpsip}. 

The weaker statement \eqref{allpsimostDp} is indeed, as Bocchieri and Loinger criticized, dynamically vacuous, as it follows straightforwardly from a statement (true for large $\dd_\nu$) that does not involve the time evolution, viz., the statement that for every $\psi$,
\be
\text{the $\decomp$-average of $\bigl(\|P_\nu\psi\|^2 -\dd_\nu/\D\bigr)^2$ is small.}
\ee
See \cite{GLMTZ09a} for a more detailed discussion.

Farquhar and Landsberg \cite{FL57} also mistook the QET for a different statement, in fact for one inequivalent to that considered by Bocchieri and Loinger. Their version differs from von Neumann's not just in the ordering of the quantifiers as in \eqref{mostDallpsip} and \eqref{allpsimostDp}, but also in that it concerns \emph{only the time average} of $\|P_\nu\psi_t\|^2$, whereas von Neumann's QET makes a statement about the value of $\|P_\nu\psi_t\|^2$ \emph{for most times}.

\section{Approach to Thermal Equilibrium}
\label{sec:mainapproach}

Von Neumann's QET, or the phenomenon of normal typicality, is closely connected with the approach to thermal equilibrium. As mentioned already, there is no consensus about what it means for a macroscopic system to be in ``thermal equilibrium.'' Before comparing the QET to more recent results in Section~\ref{sec:contemporary}, we outline in Section~\ref{sec:equilibrium} several different concepts of thermal equilibrium and in Section~\ref{sec:approach} several different concepts of approach to thermal equilibrium.

\subsection{Definitions of Thermal Equilibrium}
\label{sec:equilibrium}

We begin with the concept of thermal equilibrium that seems to us to be the most fundamental. It can be shown in many cases, and is expected to be true generally,
that for a physically reasonable choice of the macro-observables there will be among the macro-spaces $\Hilbert_\nu$ a particular macro-space $\Hilbert_{\eq}$, the one corresponding to thermal equilibrium, such that
\be\label{deqD1}
\frac{\dd_{\eq}}{\D} \approx 1\,.
\ee
In fact, the difference $1-\dd_{\eq}/\D$ is exponentially small in the number of particles. This implies, in particular, that each of the macro-observables $M_i$ is ``nearly constant'' on the energy shell $\Hilbert$ in the sense that one of its eigenvalues has multiplicity at least $\dd_{\eq}\approx \D$. 

We say that a system in the quantum state $\psi\in\sphere(\Hilbert)$ is in thermal equilibrium if and only if $\psi$ is very close (in the Hilbert space norm) to $\Hilbert_{\eq}$, or, equivalently, if and only if
\be\label{eq1}
\scp{\psi}{P_{\eq}|\psi} \approx 1\,, 
\ee
where $P_{\eq}$ is the projection operator to $\Hilbert_{\eq}$. 

The condition \eqref{eq1} implies that a quantum measurement of the macroscopic observable $M_i$ on a system with wave function $\psi$ will yield, with probability close to 1, the ``equilibrium'' value of $M_i$. Likewise, a joint measurement of $M_1,\ldots,M_\ell$ will yield, with probability close to 1, their equilibrium values.
It follows from \eqref{deqD1} that most $\psi$ on the unit sphere in $\Hilbert$ are in thermal equilibrium. Indeed, with $\mu(d\psi)$ the uniform measure on the unit sphere,
\begin{equation}\label{eq:mostVectors}
\int \bra{\psi}P_{\eq}\ket{\psi}\, \mu(d\psi) 
= \frac{\dd_{\eq}}{\D} \approx 1 \,.
\end{equation}
Since the quantity $\bra{\psi}P_{\eq}\ket{\psi}$ is bounded from above by 1, most $\psi$ must satisfy \eqref{eq1}.

If a system is normal then it is in thermal equilibrium (as defined above) most of the time.  (After all, being normal implies that $\|P_{\eq}\psi_t\|^2\approx\dd_{\eq}/\D$ most of the time, which is close to 1. Of course, if the system is not in equilibrium initially, the waiting time until it first reaches equilibrium is not specified, and may be longer than the present age of the universe.) That is why we regard the case that one of the $\Hilbert_\nu$ has the overwhelming majority of dimensions as important. Von Neumann, though, did not consider this case, and his QET actually has technical assumptions that are violated in this case. We have proved a theorem about normal typicality that applies to this case, and thus complements von Neumann's QET, in \cite{GLMTZ09b}; it asserts that for most Hamiltonians with given non-degenerate eigenvalues (or, alternatively, for most $\decomp$), all initial state vectors $\psi_0$ evolve in such a way that $\psi_t$ is in thermal equilibrium (according to the definition \eqref{eq1} above) for most times $t$.\footnote{An example of exceptional Hamiltonians that behave differently is provided by the phenomenon of \emph{Anderson localization} (see in particular \cite{And58,OH07}): Certain physically relevant Hamiltonians possess some eigenfunctions $\phi_\alpha$ that have a spatial energy density function that is macroscopically non-uniform whereas wave functions in $\Hilbert_{\eq}$ should have macroscopically uniform energy density over the entire available volume. Thus, these eigenfunctions are examples of wave functions $\psi_0$ evolving in such a way that $\psi_t$ is never in thermal equilibrium.}

The above definition of thermal equilibrium in quantum mechanics is an example of what we called the ``individualist'' view; it is analogous to the following one in classical mechanics. 
Let $\Gamma$ be the energy shell as in \eqref{Gammadef} and $\{\Gamma_\nu\}$ a partition into regions corresponding to macro-states as described in Section~\ref{sec:classical}. It has been shown \cite{Lan} for realistic systems with large $N$ that one of the regions $\Gamma_\nu$, corresponding to the macro-state of thermal equilibrium and denoted $\Gamma_{\eq}$, is such that 
\be\label{Gammaeq}
\mu_{\mc}(\Gamma_{\eq})
=\frac{\mathrm{vol}(\Gamma_{\eq})}{\mathrm{vol}(\Gamma)} 
\approx 1\,. 
\ee
We say that a classical system with phase point $X$ is in thermal equilibrium if $X\in\Gamma_{\eq}$. The analogy with the quantum mechanical definition \eqref{eq1} arises from regarding both $\psi$ and $X$ as instances of individual pure states.

\bigskip

We now turn to (what we called) the ``ensemblist'' view (for comparisons between the individualist and the ensemblist views see also \cite{Gol99,LM03}). The ensemblist defines thermal equilibrium in classical mechanics by saying that a system is in thermal equilibrium if and only if it is described by a probability distribution $\rho$ over phase space that is close to the appropriate distribution of thermal equilibrium (e.g., \cite{Tol38,Kem39,Kry,LL, Mac89,Rue08}), viz., either
\be\label{rhomuE}
\rho \approx \mu_{\mathscr{E}}\,,
\ee
where $\mu_\mathscr{E}$ is the uniform distribution on the energy surface $\Gamma_{\mathscr{E}}=\{X\in\RRR^{6N}:H(X)=\mathscr{E}\}$, or
\be
\rho \approx \mu_{\can}\,,
\ee
where $\mu_{\can}$ is the canonical distribution, which has density (relative to the phase space volume measure) proportional to
\be
e^{-\beta H(X)}\,,
\ee
with $\beta$ the inverse temperature. Correspondingly, for a quantum system the ensemblist would say that it is in thermal equilibrium if and only if it is described by a density matrix $\rho$ that is close to the appropriate density matrix of thermal equilibrium (e.g., \cite{Tol38,Kem39,LL}), viz., either
\be\label{rhomc}
\rho\approx \rho_{\mc}
\ee
or
\be\label{rhocan}
\rho\approx\rho_{\can}\,,
\ee
where $\rho_{\can}$ is the canonical density matrix
\be\label{candef}
\rho_{\can} = \frac{1}{Z} e^{-\beta H}\,,\quad
Z=\tr e^{-\beta H}\,,
\ee
and $H$ is the system's Hamiltonian.

When considering a single classical system, the individualist insists that it has a phase point $X$, well-defined though usually unknown, but no distribution $\rho$ (except the delta distribution $\delta_X$) because it is a single system; the ensemblist thinks that our knowledge of $X$, which is always incomplete, should be represented by a probability distribution $\rho$. Thus, the individualist regards thermal equilibrium as an objective event, the ensemblist as a subjective one---his notion of thermal equilibrium has an information-theoretic nature. 

Other definitions of thermal equilibrium are inspired by both the ensemblist and the individualist views. Von Neumann would have said, we think, that a system with wave function $\psi\in\sphere(\Hilbert)$ is in thermal equilibrium if and only if $\|P_\nu \psi\|^2 \approx \dd_\nu/\D$ for all $\nu$, i.e., if the probability distribution over the $\nu$s defined by $\psi$ coincides approximately with the micro-canonical distribution, or $\ket{\psi}\bra{\psi}\stackrel{\decomp}{\sim}\rho_{\mc}$. This definition has in common with the individualist definition \eqref{eq1} (and differs from the ensemblist definition \eqref{rhomc} or \eqref{rhocan} in) that it can be satisfied for a system in a pure state. The ensemblist spirit comes into play when considering whether the \emph{probability distribution} of $\nu$ is close to micro-canonical; while in classical mechanics, an individual phase point $X$ defines only a delta distribution, which is far from micro-canonical, in quantum mechanics $\psi$ defines a distribution over $\nu$ that is indeed approximately micro-canonical, even for an individual system. 

That von Neumann defined thermal equilibrium in this way explains why he did not consider the case that one of the $\dd_\nu$s is close to $\D$. Note also that, in terms of this definition, being normal immediately means being in thermal equilibrium most of the time, so that the QET is a statement about thermal equilibrium (although von Neumann never explicitly mentioned thermal equilibrium in his QET article). 

Here is another definition inspired by both views. Consider a bi-partite system consisting of subsystem 1 and subsystem 2, with Hilbert space $\Hilbert_{1\cup 2} = \Hilbert_1 \otimes \Hilbert_2$, and suppose it is in a pure state $\psi\in\Hilbert_{1\cup 2}$. One might say that subsystem 1 is in thermal equilibrium if and only if 
\be\label{rho1can}
\rho_1 \approx \rho_{\can}\,,
\ee
where $\rho_1 = \tr_2 \pr{\psi}$ is the reduced density matrix of subsystem 1, and $\rho_{\can}$ is given by \eqref{candef} with some $\beta$ and the Hamiltonian of subsystem 1 in the place of $H$. (Also, one might say that a system is in thermal equilibrium if every small subsystem has reduced density matrix that is approximately canonical.) While the ensemblist spirit is visible in the similarity between \eqref{rhocan} and \eqref{rho1can}, this definition is more on the individualist side because the whole system is assumed to be in a pure state.

\subsection{Definitions of Approach to Thermal Equilibrium}
\label{sec:approach}

Corresponding to the different notions of what it means for a quantum system to be in thermal equilibrium, there are different notions of what it means to \emph{approach} thermal equilibrium. Individualists (like us) consider an isolated system of finitely many particles in a pure state $\psi$ whose time evolution is unitary and say that the system approaches thermal equilibrium if and only if $\psi_t$, for some $t>0$, belongs to the set of $\psi$s in thermal equilibrium (and remains in that set for a very long time). The ensemblist, one might imagine, would say that a system approaches thermal equilibrium if and only if its density matrix $\rho_t$, for some $t>0$, is close to the appropriate density matrix of thermal equilibrium ($\rho_{\mc}$ or $\rho_{\can}$). However, inspired by the situation in classical mechanics (see below), ensemblists tend to demand more and to say that a system approaches thermal equilibrium if and only if its density matrix $\rho_t$ converges, as $t\to\infty$, to $\rho_{\mc}$ or $\rho_{\can}$.

This ensemblist notion of approach to thermal equilibrium is certainly mathematically appealing. However, it is very hard for it to hold: Consider an isolated system of finitely many particles with unitary time evolution. Then neither a (non-equilibrium) mixed state $\rho_t$ of that system nor the reduced density matrix $\rho_{1,t}=\tr_2 \rho_t$ of a subsystem will converge, as $t\to\infty$, to $\rho_{\mc}$ or $\rho_{\can}$. This is because of the \emph{recurrence} properties of the unitary evolution: If the Hilbert space is finite-dimensional (which is the case if we consider only a finite energy interval such as $[\mathscr{E}_a,\mathscr{E}_{a+1})$ for finitely many particles in a finite volume) then there are arbitrarily large $t>0$ such that the unitary time evolution operator
\be
U_t=\exp(-iHt/\hbar)
\ee
is arbitrarily close to the identity operator. This fact is a consequence of the \emph{quasi-periodicity} of the unitary evolution. Thus, the density matrix
\be
\rho_t=U_t \, \rho_0 \, U_t^*
\ee
keeps on returning to near its initial state, and so does $\rho_{1,t} = \tr_2 \rho_t$.\footnote{If $\Hilbert$ is infinite-dimensional and $H$ has discrete spectrum (as it would if the system is confined to a finite volume), then $U_t$ may not be close to the identity for any $t>0$, but still every density matrix $\rho_0$ keeps on returning to near its initial state. Indeed, $\rho_0$ can be approximated by an operator $\rho_0'$ of finite rank (i.e., a mixture of only finitely many pure states), which in turn can be approximated by an operator $\rho''_0$ on a subspace spanned by finitely many energy eigenstates, and the quasi-periodicity implies that there are arbitrarily large $t>0$ with $\rho_t''\approx \rho_0''$; at such times, also $\rho_t\approx \rho_0$.}

It also follows from recurrence that for an individualist, a system starting with a non-equilibrium state $\psi_0$ cannot remain forever in thermal equilibrium after reaching it. A valid statement can assert at best that the system will spend most of the time in thermal equilibrium; that is, it will again and again undergo excursions away from thermal equilibrium, but in between spend overwhelmingly long periods in thermal equilibrium.
In fact, our theorem in \cite{GLMTZ09b} asserts that when $\D$ is large enough and one $\dd_\nu=\dd_{\eq}$ has the vast majority of dimensions as in \eqref{deqD1} then, for most non-degenerate Hamiltonians, all initial pure states $\psi_0$ will spend most of the time in thermal equilibrium in the sense of the definition \eqref{eq1}.

To avoid recurrence in quantum physics, the ensemblist considers infinite systems (which cannot be described by means of Hilbert spaces but by $C^*$ or $W^*$ algebras), for example a finite system of interest coupled to an infinitely large heat bath that is initially in thermal equilibrium (see, e.g., \cite{vH57,LS78}). In fact, for such situations, the convergence $\rho_{1,t} \to\rho_{\can}$ has been proved rigorously \cite{R73,JP,BFS00}; see also \cite{VH09} and references therein. Thus, for the ensemblist the approach to thermal equilibrium is an idealization and never occurs in the real world.

Curiously, the ensemblist's invocation of infinite systems is unnecessary in classical mechanics because the time evolution of a classical system is usually not quasi-periodic, and the recurrence of mixed states described above need not arise. The ensemblist's approach to thermal equilibrium is then connected to the property of being \emph{mixing} \cite{Kry}, a property related to (but stronger than) ergodicity: For any probability distribution $\rho_0$ of $X_0$, let $\rho_t$ denote the distribution of $X_t$. The dynamics on $\Gamma_{\mathscr{E}}$ is mixing if, for every absolutely continuous probability distribution $\rho_0$ (i.e., one that has a density relative to $\mu_\mathscr{E}$), $\rho_t\to\mu_\mathscr{E}$ as $t\to\infty$ (in the sense of convergence on bounded functions). 
In contrast, the individualist expects that for realistic classical systems with a sufficiently large number $N$ of constituents and for every macro-state $\nu$, most initial phase points $X\in\Gamma_\nu$ will be such that $X_t$ spends most of the time in the set $\Gamma_{\eq}$. This statement follows if the time evolution in phase space is ergodic, but in fact is much weaker than ergodicity.

\subsection{Current Research on the Approach to Thermal Equilibrium}
\label{sec:contemporary}

Various results about the approach to equilibrium in the individualist framework have been obtained in recent years \cite{Deu91,Sre94,T98,R08,RDO08, LPSW08,GLMTZ09b}. Many of these results can be described in a unified way as follows. Let us say that a system with initial wave function $\psi(0)$ \emph{equilibrates} relative to a class $\A$ of observables if for most times $\tau$,
\be\label{equidef}
\scp{\psi_{\tau}}{A|\psi_{\tau}} \approx 
\tr\Bigl(\overline{\ket{\psi_t}\bra{\psi_t}}A\Bigr) 
\text{ for all }A\in\A\,.
\ee
We then say that the system \emph{thermalizes} relative to $\A$ if it equilibrates and, moreover,
\be
\tr\Bigl(\overline{\ket{\psi_t}\bra{\psi_t}} A\Bigr)\approx
\tr\bigl(\rho_{\mc}A\bigr) \text{ for all }A\in\A\,.
\ee
That is, the system thermalizes relative to $\A$ if, for most times $t$,
\be\label{thermdef}
\scp{\psi_t}{A|\psi_t} \approx 
\tr\bigl(\rho_{\mc}A\bigr)
\text{ for all }A\in\A\,.
\ee
With these definitions, the results of \cite{T98,R08,LPSW08,GLMTZ09b}, as well as von Neumann's QET, can be formulated by saying that, under suitable hypotheses on $H$ and $\psi(0)$ and for large enough $\D$, a system will equilibrate, or even thermalize, relative to a suitable class $\A$. (It should in fact be true for a large class of observables $A$ on $\Hilbert$ that, for most $\psi$ on the unit sphere in $\Hilbert$, $\scp{\psi}{A|\psi} \approx \tr(\rho_{\mc}A)$; if this is true of every member of $\A$ then it is not hard to see that most initial wave functions will thermalize relative to $\A$.)

Von Neumann established in his QET, under assumptions we described in Section~\ref{sec:qualitative}, thermalization for a family $\A$ of commuting observables; $\A$ is the algebra generated by $\{M_1,\ldots,M_\ell\}$. Rigol, Dunjko, and Olshanii \cite{RDO08} numerically simulated a model system and concluded that it thermalizes relative to a certain class $\A$ consisting of commuting observables. Our result in \cite{GLMTZ09b} takes $\A$ to contain just one operator, namely $P_{\eq}$. We established thermalization for arbitrary $\psi(0)$ assuming $H$ is non-degenerate and satisfies $\scp{\phi_\alpha}{P_{\eq}|\phi_\alpha}\approx 1$ for all $\alpha$, which (we showed) is typically true. Tasaki \cite{T98} as well as Linden, Popescu, Short, and Winter \cite{LPSW08} considered a system coupled to a heat bath, $\Hilbert_\mathrm{total}=\Hilbert_\mathrm{sys}\otimes\Hilbert_\mathrm{bath}$, and took $\A$ to contain all operators of the form $A_\mathrm{sys}\otimes 1_\mathrm{bath}$. Tasaki considered a rather special class of Hamiltonians and established thermalization assuming that many eigenstates of $H$ contribute to $\psi_0$. Under a similar assumption on $\psi_0$, Linden et al.\ established equilibration for $H$ without resonances. They also established a result in the direction of thermalization under the additional hypothesis that the dimension of the energy shell of the bath is much greater than $\dim \Hilbert_\mathrm{sys}$. Reimann's mathematical result \cite{R08} can be described in the above scheme as follows. Let $\A$ be the set of all observables $A$ with (possibly degenerate) eigenvalues between 0 and 1 such that the absolute difference between any two eigenvalues is at least (say) $10^{-1000}$. He established equilibration for $H$ without resonances, assuming that many eigenstates of $H$ contribute to $\psi_0$.

\section{The Method of  Appeal to Typicality}
\label{sec:typicality}

We would like to clarify the status of statements about ``most'' or ``typical'' $\decomp$ (or, for that matter, most $H$ or most $\psi_0$), and in so doing elaborate on von Neumann's method of appeal to typicality. In 1955, Fierz criticized this method as follows \cite[p.~711]{F55}:
\begin{quotation}
The physical justification of the hypothesis [that all $\decomp$s are equally probable] is of course questionable, as the assumption of equal probability for all observers is entirely without reason. Not every macroscopic observable in the sense of von Neumann will really be measurable. Moreover, the observer will try to measure exactly those quantities which appear characteristic of a given system. 
\end{quotation}
In the same vein, Pauli  wrote 
in a private letter to Fierz in 1956 \cite{P56}:
\begin{quotation}
As far as assumption B [that all $\decomp$s are equally probable] is concerned [\ldots] I consider it \emph{now} not only as lacking in plausibility, but \emph{nonsense}. 
\end{quotation}
Concerning these objections, we first note that it is surely
informative that normality holds for some $\decomp$s, let alone that it
holds in fact for most $\decomp$s, with ``most'' understood in a
mathematically natural way. But we believe that more should be said. 

When  employing the method of appeal to typicality, one usually uses the language of probability theory. But that does not imply that any of the objects considered is random in reality. Rather, it means that certain sets (of wave functions, of orthonormal bases, etc.)\ have certain sizes (e.g., close to 1) in terms of certain natural (normalized) measures of size. That is, one describes the behavior that is \emph{typical} of wave functions, orthonormal bases, etc.. However, since the mathematics is equivalent to that of probability theory, it is convenient to adopt that language. For this reason, using a normalized measure $\mu$ does not mean making an ``assumption of equal probability,'' even if one uses the word ``probability.'' Rather, it means that, if a condition is true of most $\decomp$, or most $H$, this fact may suggest that the condition is also true of a concrete given system, unless we have reasons to expect otherwise. 

Of course, a theorem saying that a condition is true of the vast majority of systems does not \emph{prove} anything about a concrete given system; if we want to know for sure whether a given system is normal for every initial wave function, we need to check the relevant condition, which is \eqref{cond2a} above. Nevertheless, a typicality theorem is, as we have suggested, illuminating; at the very least, it is certainly useful to know which behaviour is typical and which is exceptional. Note also that the terminology of calling a system ``typical'' or ``atypical'' might easily lead us to wrongly conclude that an  atypical system  will not be normal. A given system may have some properties that are atypical and nevertheless satisfy the condition \eqref{cond2a} implying that the system is normal for every initial wave function.

The method of appeal to typicality belongs to a long tradition in physics, which includes also Wigner's work on random matrices of the 1950s. In the words of  
Wigner \cite{Wigner}:
\begin{quote}
One [\dots] deals with a specific system, 
with its proper (though in many cases unknown) Hamiltonian, yet pretends 
that one deals with a multitude of systems, all with their own Hamiltonians, 
and averages over the properties of these systems. Evidently, such a procedure 
can be meaningful only if it turns out that the properties in which one is interested are the same for the vast majority of the admissible Hamiltonians.
\end{quote}
This method was used by Wigner to obtain specific new and surprising predictions about detailed properties of complex quantum systems in nuclear physics. 

If we know of a given system that its Hamiltonian $H$ belongs to a particular small subset $S_0$ of the set $S$ of all self-adjoint operators on the appropriate Hilbert space, then two kinds of typicality theorems are of interest: one saying that the relevant behavior occurs for most $H\in S_0$, the other saying that it occurs for most $H\in S$. Note that the former does not follow from the latter when $S_0$ is very small compared to $S$, as it would then be consistent with the latter for $S_0$ to consist exclusively of exceptional $H$s. Nor does the latter follow from the former, so the two statements are logically independent. In fact, both are of interest because each statement has its merits: The typicality theorem about $S_0$ gives us more certainty that the given system, whose Hamiltonian belongs to $S_0$, will behave in the relevant way. The typicality theorem about $S$ gives us a deeper understanding of \emph{why} the relevant behavior occurs, as it indicates that the behavior has not much to do with $S_0$ but is widespread all over $S$. That is, there is a reciprocal relation: The greater the degree of certainty that a typicality theorem confers, the less its explanatory power.

\section{Von Neumann's Quantum $H$-Theorem}
\label{sec:mainHtheorem}

In his proof of the QET, von Neumann describes parallel considerations that prove a second theorem that he calls the ``quantum $H$-theorem.'' This concerns the long-time behavior of the quantity $S$ that von Neumann defines in his equation (34) to be the entropy of a system with wave function $\psi$:
\be\label{Sdef}
S(\psi) =-k\sum_\nu \|P_\nu \psi\|^2 \log \frac{\|P_\nu \psi\|^2}{\dd_\nu}\,,
\ee
where $k$ is the Boltzmann constant and $\log$ denotes the natural logarithm. (This formula looks a bit simpler than von Neumann's (34) because we consider only $\psi$s that lie in a particular micro-canonical Hilbert space $\Hilbert$.) Note that this definition is different from the one usually known as the \emph{von Neumann entropy},
\be\label{SvN}
S_{\mathrm{vN}}(\rho) = -k \tr(\rho \log \rho)
\ee
for a system with density matrix $\rho$, which had been introduced by von Neumann two years earlier \cite{vN27c}. 

In Section~\ref{sec:defentropy} we discuss the definition \eqref{Sdef}. As with thermal equilibrium, there is no consensus about the definition of entropy in quantum mechanics. In Section~\ref{sec:Htheorem} we discuss the contents of von Neumann's ``quantum $H$-theorem.''

\subsection{Von Neumann's Definition of Entropy}
\label{sec:defentropy}

We begin by giving a brief overview of several approaches. We first recall Boltzmann's entropy definition \cite{B96} for a macroscopic classical system. Consider an energy shell $\Gamma$ as defined in \eqref{Gammadef}, partitioned into subsets $\Gamma_\nu$ corresponding to different macro-states $\nu$. A system with phase point $X\in\Gamma$ has \emph{Boltzmann entropy}
\be\label{SBoltzmann}
S_{\mathrm{B}}(X) = k\log \mathrm{vol}(\Gamma_\nu) \quad 
\text{if and only if}
\quad X\in \Gamma_\nu\,,
\ee
where $\mathrm{vol}(\Gamma_\nu)$ is the phase space volume of $\Gamma_\nu$; see also \cite{Gol99,L99,L07}. 
As a quantum mechanical analog, consider a macroscopic quantum system, an energy shell $\Hilbert$ as defined in Section~\ref{sec:setting}, and an orthogonal decomposition into subspaces $\Hilbert_\nu\subseteq \Hilbert$ corresponding to different macro-states $\nu$; define the \emph{quantum Boltzmann entropy} of a system with wave function $\psi\in\Hilbert_\nu$ by
\be\label{Snu}
S_{\mathrm{qB}}(\psi) = k\log \dd_\nu\,.
\ee
We also denote this quantity by $S_\mathrm{qB}(\nu)$. (See also \cite{L07} for a discussion of this formula. A version of it, with $\dd_\nu$ the ``number of elementary states,'' was used already by Einstein in 1914 \cite[Eq.~(4a)]{Ein14}.)

While the Boltzmann entropy is based on the individualist view, there is a counterpart in the ensemblist view, known as the \emph{Gibbs entropy} in classical mechanics; it is defined for a system described by the probability density $\rho$ on phase space $\RRR^{6N}$ to be
\be\label{SGibbs}
S_{\mathrm{G}}(\rho) = -k \int_{\RRR^{6N}} dX\, \rho(X) \, \log \rho(X)\,.
\ee
Its quantum mechanical analog is the von Neumann entropy, defined for a system described by the density matrix $\rho$ on $\Hilbert_{\total}$ by \eqref{SvN}. In his QET article, von Neumann writes in Section 1.3: ``The [expression \eqref{SvN}] for entropy 
[... is] not applicable here in the way [it was] intended, as [it was] computed from the perspective of an observer who can carry out all measurements that are possible in principle---i.e., regardless of whether they are macroscopic.'' 
We agree that \eqref{SvN} is not applicable to the macroscopic quantum system von Neumann considers in his QET article, but for a different reason. Von Neumann's reason is of an ensemblist, information-theoretic nature, supposing that the value of the entropy quantifies the (possible) knowledge of an observer. In an individualist framework, where the system is regarded as being in a pure state, the von Neumann entropy \eqref{SvN} is clearly inadequate because it always yields the value zero; more fundamentally, an individualist regards entropy not as measuring the spread of a probability distribution, but as measuring the size of a macro-state. However, note that
for $\rho = \dd_\nu^{-1}P_\nu$ and $\psi\in\Hilbert_\nu$,
\be
S_\mathrm{qB}(\psi) = S_\mathrm{vN}(\rho)\,.
\ee

Let us turn to a comparison between \eqref{Sdef} and \eqref{Snu}. Since, for a macroscopic system, the contribution $k\sum_\nu \|P_\nu\psi\|^2 \log \|P_\nu\psi\|^2$ is sufficiently small \cite{F55}, we have that
\be\label{Spsisnu}
S(\psi) \approx k\sum_\nu \|P_\nu \psi\|^2 \log \dd_\nu\,.
\ee
This quantity is just the weighted average of the $S_{\mathrm{qB}}(\nu)$, with the weight of $\nu$ given by the quantum-mechanical probability of $\nu$ associated with $\psi$. 

We conclude from this relation that $S(\psi)$ should better be regarded as a sort of \emph{mean entropy} of the system, than as its \emph{entropy}. For comparison, for a classical system whose macro-state is unknown and has probability $p_\nu$ to be $\nu$, we would not say that the quantity
\be\label{averageS}
\sum_\nu p_\nu \, k \log \mathrm{vol}(\Gamma_\nu)\,,
\ee
is the entropy of the system, but we would say instead that the system's entropy $S$ is random, that it equals $k\log\mathrm{vol}(\Gamma_\nu)$ with probability $p_\nu$, and that \eqref{averageS} is its expected value $\EEE S$. Now, the quantum situation is not completely analogous because a quantum superposition of contributions from different $\Hilbert_\nu$ is not the same as a statistical mixture of wave functions from different $\Hilbert_\nu$, but the analogy is good enough to make us doubt the adequacy of von Neumann's definition \eqref{Sdef}. 

Moreover, there are situations in which $S(\psi)$ decreases, contrary to the second law of thermodynamics. If $\psi$ is a non-trivial superposition
\be\label{superposition}
\psi = \sum_\nu c_\nu \psi_\nu
\ee
of macroscopically different quantum states $\psi_\nu\in\Hilbert_\nu$ with $\|\psi_\nu\|=1$ (such as, e.g., a Schr\"odinger cat state) then a measurement of all macro-observables will yield, with probability $|c_\nu|^2$, the macro-state $\nu$ and the wave function $\psi_\nu$ with $S(\psi_\nu)= S_{\mathrm{qB}}(\psi_\nu)$. Since $S(\psi)$, in the approximation \eqref{Spsisnu}, is just the average of the random value $S(\psi_\nu)$, the $S(\psi_\nu)$ can (and usually will) have significant probability (possibly 50\%\ or even more) to lie below $S(\psi)$.

\subsection{Von Neumann's Quantum $H$-Theorem}
\label{sec:Htheorem}

In the previous subsection, we have expressed reservations about the adequacy of von Neumann's definition $S(\psi)$ of the entropy of a macroscopic quantum system. Be that as it may, here is what von Neumann's quantum $H$-theorem asserts (roughly): 
\textit{For any $H$ without resonances on $\Hilbert$, most orthogonal decompositions $\decomp$ of $\Hilbert$ are such that for every wave function $\psi_0$, $S(\psi_t)$ will be close to its upper bound $k\log\D$ for most times $t$.} (The detailed statement in the QET article specifies a bound on how close, see von Neumann's Equations (70) and (79)--(84).)

We would like to emphasize that this statement is really just a corollary of the QET. If there is a subspace $\Hilbert_{\eq}$ with $\dd_{\eq}/\D\approx 1$ then for any wave function $\psi$ in thermal equilibrium as defined in \eqref{eq1}, \eqref{Spsisnu} entails that
\be
S(\psi)\approx k\log \D\,.
\ee
Thus, the qualitative content of the above statement follows already if the system is in thermal equilibrium for most of the time, and thus follows from normal typicality. More generally, even if there is no equilibrium macro-state obeying $\dd_{\eq}/\D\approx 1$, the statement follows from \eqref{Spsisnu} and the QET, together with the condition that the number $n$ of macro-states $\nu$ be not too large. (Indeed, for most $t$, according to the QET, then
\be\label{expr2}
S(\psi_t) \approx k \sum_\nu \frac{\dd_\nu}{\D} \log\dd_\nu\,.
\ee
For those $\nu$ with (say) $\dd_\nu \geq \D/10^3n$, we have that $\log \dd_\nu \geq \log \D - \log(10^3n) \approx \log \D$ if $n$ is not too large; furthermore, the sum of these $\dd_\nu$s is at least $0.999 \, \D$ because the sum of the remaining $\dd_\nu$s is a sum of less than $n$ terms, each of which is less than $\D/10^3n$, and thus is less than $\D/10^3$. Thus, the right hand side of \eqref{expr2} is at least $0.999\, k (\log \D - \log(10^3n))$, which is approximately $k\log D$.)

More importantly, there are striking differences between von Neumann's quantum $H$-theorem and Boltzmann's classical $H$-theorem \cite{B96} (see also \cite{Lan75,Gol99,L99,GL}). Boltzmann's $H$-theorem was originally formulated only for systems whose behavior is well described by the Boltzmann equation (i.e., dilute, weakly interacting gases) but can be understood in a more general sense as the assertion that \textit{for most initial phase points $X_0\in \Gamma_\nu\neq \Gamma_{\eq}$, the Boltzmann entropy $S_{\mathrm{B}}(X_t)$ increases, up to exceptions, monotonically in both time directions with $|t|$ until it reaches the maximal (equilibrium) value; the exceptions (entropy valleys) are either very short-lived and shallow, or infrequent.} This statement is not a mathematical theorem but very plausible.\footnote{Von Neumann apparently did not think of this statement when thinking of the classical $H$-theorem. He wrote in Section 0.6: ``As in classical mechanics, also here there is no way that entropy could always increase, or even have a predominantly positive sign of its [time] derivative (or difference quotient): the time reversal objection as well as the recurrence objection are valid in quantum mechanics as well as in classical mechanics.'' In fact, the above statement of the $H$-theorem conveys something that could be called a ``predominantly positive sign'' of the time difference quotient of $S=S_{\mathrm{B}}(X_t)$ but is not refuted by either time reversal or recurrence. Put very succinctly, this is because, according to the statement, $S$ increases in both time directions and may well decrease after reaching its maximum; see, e.g., \cite{Gol99,L99,GL} for further discussion.}

In contrast, what von Neumann's quantum $H$-theorem implies about the increase of $S(\psi_t)$ (for a system without resonances in the Hamiltonian and a typical decomposition $\decomp$) is much less: \textit{If, initially, $S(\psi_0)$ is far below $k\log \D$ then it will in both time directions sooner or later reach the maximal (equilibrium) value $k\log\D$, i.e., there are $t_+>0$ and $t_-<0$ such that $S(\psi_{t_+})\approx k\log \D \approx S(\psi_{t_-})$; moreover, $S(\psi_t)$ will assume its maximal value for most $t$ in the long run (in both time directions).} However, no statement is implied about a largely \emph{monotone} increase; for example, insofar as von Neumann's quantum $H$-theorem is concerned, $S(\psi_t)$ may first go down considerably before increasing to $k\log \D$. Likewise, no statement is made about the features of the entropy valleys between $0$ and $t_\pm$; they could, perhaps even for most wave functions, be long-lived, deep, and frequent.

\appendix

\section{Precise Statement of von Neumann's Quantum Ergodic Theorem}
\label{sec:thm}

In von Neumann's article, the statement of the quantum ergodic theorem is distributed over many pages. It may therefore be helpful for the reader if we provide the exact statement of the QET. Readers interested in the exact statement may also wish to look at the two modified versions of the statement that we have described in \cite{GLMTZ09a}; although they are in some ways stronger, they also follow from von Neumann's proof.

\begin{defn}
The system corresponding to a choice of $\Hilbert$, $\psi_0\in\sphere(\Hilbert)$, $H$, and $\decomp$ is $\varepsilon$-$\delta'$-normal if and only if, for $(1-\delta')$-most $t$,
\be\label{vNdeforig}
\bigl|\scp{\psi_t}{A|\psi_t} - \tr(\rho_{\mc} A) \bigr|< \varepsilon \sqrt{\tr(\rho_{\mc} A^2)}
\ee
for every real-linear combination $A=\sum_\nu \alpha_\nu P_\nu$ (i.e., for every self-adjoint operator $A$ from the algebra generated by the macro-observables $M_1,\ldots,M_\ell$).
\end{defn}

The condition \eqref{vNdeforig} is more or less equivalent, when the number $n$ of macro-spaces $\Hilbert_\nu$ in $\Hilbert$ is much smaller than each of the $\D/\dd_\nu$, to the condition that
\be\label{vNdef}
\Bigl|\|P_\nu\psi_t\|^2 - \frac{\dd_\nu}{\D} \Bigr|< \varepsilon \sqrt{\frac{\dd_\nu}{n\D}}
\quad \text{for all }\nu\,,
\ee
which is a precise version of \eqref{approx}. See Section 4 of \cite{GLMTZ09a} for more details about the relation between \eqref{vNdeforig} and \eqref{vNdef}.

\begin{thm}\label{thm:vN}
(von Neumann's 1929 QET)
Let $\varepsilon>0$, $\delta>0$, and $\delta'>0$. 
For arbitrary $\Hilbert$ of finite dimension, any orthogonal decomposition $\decomp=\{\Hilbert_\nu\}$ of $\Hilbert$, and any $H$ without resonances, if
\be\label{cond2b}
\max_{\alpha\neq \beta}
\bigl|\scp{\phi_\alpha}{P_\nu|\phi_\beta}\bigr|^2 
+ \max_\alpha 
\Bigl(\scp{\phi_\alpha}{P_\nu|\phi_\alpha}-\frac{\dd_\nu}{\D}\Bigr)^2
< \varepsilon^2 \frac{\dd_\nu}{n\D}\frac{\delta'}{n} 
\quad \text{for all }\nu
\ee
with $\dd_\nu=\dim\Hilbert_\nu$, $\D=\dim\Hilbert$, and $n=\#\decomp$, 
then for every $\psi_0\in\sphere(\Hilbert)$ the system is $\varepsilon$-$\delta'$-normal. Moreover, suppose we are given natural numbers $\D$, $n$, and $\dd_1,\ldots,\dd_n$ such that $\dd_1+ \ldots +\dd_n=\D$ and, for all $\nu$,
\be\label{cond5}
\max\Bigl(C_1, \frac{10n^2}{\varepsilon^2\delta'\delta} \Bigr) \log \D < \dd_\nu < \D/C_1\,,
\ee
where $C_1$ is a universal constant. Then $(1-\delta)$-most orthogonal decompositions $\decomp=\{\Hilbert_\nu\}$ of $\Hilbert$ with $\dim\Hilbert_\nu = \dd_\nu$ are such that \eqref{cond2b} is satisfied (and thus the system is $\varepsilon$-$\delta'$-normal for every $\psi_0\in\sphere(\Hilbert)$).
\end{thm}

One of the stronger versions of the QET described in \cite{GLMTZ09a} asserts that, under somewhat stronger assumptions on the $\dd_\nu$s, \eqref{vNdeforig} and \eqref{vNdef} can be replaced with the stronger (and more natural) error bound
\be\label{strongdef}
\Bigl|\|P_\nu\psi_t\|^2 - \frac{\dd_\nu}{\D} \Bigr|< \varepsilon \frac{\dd_\nu}{\D}
\quad \text{for all }\nu\,.
\ee

\section{Von Neumann's Notation and Terminology}
\label{sec:notation}

The following table lists some of von Neumann's notation, and the (different) notation we use here.

\begin{center}
\begin{tabular}{ccl}
vN & here & meaning\\\hline
$\Aa,\Bb,\ldots$ & $A,B,\ldots$ & operators on Hilbert space\\
$(\phi,\psi)$ & $\langle\psi|\phi\rangle$ & inner product in Hilbert space\\
$\Hh$ & $H$ & Hamiltonian operator\\
$\varphi_{\rho,a}, W_{\rho,a}$ & $\phi_\alpha,E_\alpha$ & eigenfunction, eigenvalue of $\Hh$\\
$\Pp_\psi$ & $|\psi\rangle\langle\psi|$ & projection to 1-d subspace spanned by $\psi$\\
$\Uu$ & $\rho$ & density operator\\
& $\mathscr{I}_a=[\mathscr{E}_a,\mathscr{E}_{a+1})$ & $a$-th energy interval \\
& $\Hilbert_{\mathscr{I}_a} = \Hilbert$ & $a$-th energy shell\\
$\Ddelta_a$ & & projection to the $a$-th energy shell\\
$S_a$ & $\D$ & dimension of the $a$-th energy shell\\
$N_a$ & $n$ & number of macro-spaces in the $a$-th energy shell\\
& $\Hilbert_\nu$ & $\nu$-th macro-space (in the $a$-th energy shell)\\
$\Ee_{\nu,a}$ & $P_\nu$ & projection to $\Hilbert_\nu$\\
$s_{\nu,a}$ & $\dd_\nu$ & dimension of $\Hilbert_\nu$\\
$\omega_{\lambda,\nu,a}$ & & an orthonormal basis of $\Hilbert_\nu$\\
$M_t\{f(t)\}$ & $\overline{f(t)}$ & time average of $f(t)$ \\
$\M$ & $\EEE$ & ensemble average
\end{tabular}
\end{center}

Since for us, but not for von Neumann, the index $a$ is fixed throughout, it is usually omitted in our notation.

When von Neumann spoke of an \emph{orthogonal system} in Hilbert space in the QET article, he meant an ortho\emph{normal} system; this becomes clear from his Equation (20) in Footnote 25, where he uses $\Pp_\varphi f = (f,\varphi)\varphi$, which is true only when $\|\varphi\|=1$, for an element $\varphi$ of an ``orthogonal system.''

Von Neumann used the expression \emph{macroscopic observer} in the QET article when referring to the family $\decomp=\{\Hilbert_\nu\}$ of macro-spaces, or, equivalently, to the family $\{M_1,\ldots,M_\ell\}$ of commuting observables. This is perhaps not a fitting terminology, as it may suggest that different people would have different sets like $\{M_1,\ldots,M_\ell\}$ associated with them, which is not the case. It is perhaps even an unfortunate terminology, as it may further suggest that the (uniform) distribution over all $\decomp$s that the QET represents the distribution of different people's $\decomp$ in some population of observers. Instead, this distribution should be regarded as just the mathematical means for expressing what is true of most orthogonal decompositions $\decomp$.

Von Neumann used the expression \emph{energy surface} in quantum mechanics in different sections of his article with different meanings. The main meaning, used in his Section 1, is what we call an energy shell: The subspace $\Hilbert$ corresponding to a narrow interval $[\mathscr{E}_a,\mathscr{E}_{a+1})$ of energies. (Sometimes, he used ``energy surface'' when referring to the corresponding micro-canonical density matrix $\rho_{\mc}$, which is $1/\D$ times the projection to $\Hilbert$, or, in von Neumann's notation, $(1/S_a)\Ddelta_a$.) In Section 0.4, he used ``energy surface'' for the torus \eqref{torus} with fixed radii $r_\alpha\geq 0$.

The expression \emph{micro-canonical ensemble} undergoes a similar change in meaning: While in Section 0.4, it refers to the uniform distribution over the torus \eqref{torus}, its main meaning, used in Section 1, is what we call the micro-canonical density matrix $\rho_{\mc}$. However, besides that, von Neumann also used that expression for another density matrix $\Uu_\psi$, associated with a given wave function $\psi\in\Hilbert_{\total}$ with contributions from several different energy shells: it is a mixture of the projections to the different energy shells $\Hilbert_{\mathscr{I}_a}$ with weights provided by $\psi$. In his words (after his Equation (32) in Section 1.3): ``Now we are ready to define the micro-canonical ensemble pertaining to the state $\psi$. [... W]e define it to be the mixture of the $\frac{1}{S_1}\Ddelta_1,\frac{1}{S_2}\Ddelta_2,\ldots$ with weights $(\Ddelta_1\psi,\psi), (\Ddelta_2\psi,\psi),\ldots$.'' This further complication vanishes when one focuses, as one can without damage, on wave functions from one particular energy shell.

\bigskip

\noindent\textit{Acknowledgements.}
We thank Wolf Beiglb\"ock for helpful remarks. 
S.~Goldstein was supported in part by National Science Foundation [grant DMS-0504504].
N.~Zangh\`\i\ was supported in part by Istituto Nazionale di Fisica Nucleare. 
J.~L.~Lebowitz was supported in part by NSF [grant DMR 08-02120] and by AFOSR [grant AF-FA 09550-07].

\end{document}